%% file: arxiv.tex
\documentclass{article}

\usepackage{arxiv}

\usepackage{threeparttable}
\usepackage{tabularx}
\usepackage{amsmath}
\usepackage{enumerate}

\usepackage[utf8]{inputenc} 
\usepackage[T1]{fontenc}    
\usepackage{hyperref}       
\usepackage{url}            
\usepackage{booktabs}       
\usepackage{amsfonts}       
\usepackage{nicefrac}       
\usepackage{microtype}      
\usepackage{cleveref}       
\usepackage{lipsum}         
\usepackage{graphicx}
\usepackage{natbib}
\usepackage{doi}

\usepackage[group-separator={,},output-decimal-marker={.}]{siunitx}

\newcommand{\mgiven}{\,|\,}

\title{Prediction can be safely used as a proxy for explanation in causally consistent Bayesian generalized linear models}

\date{}

\author{
Maximilian Scholz \\
	Cluster of Excellence SimTech\\
	University of Stuttgart\\
 Germany \\
	\texttt{research.scholz@mailbox.org} \\
	\And
 Paul-Christian Bürkner \\
	Department of Statistics\\
 TU Dortmund University\\
 Germany\\
}

\renewcommand{\headeright}{}
\renewcommand{\shorttitle}{Prediction as proxy for explanation}

\hypersetup{
pdftitle={Prediction can be safely used as a proxy for explanation in causally consistent Bayesian generalized linear models},
pdfauthor={Maximilian Scholz, Paul-Christian Bürkner},
pdfkeywords={Bayesian workflow, causal inference, explanation, prediction, generalized linear models, simulation study},
}

\begin{document}
\maketitle

\begin{abstract}
Bayesian modeling provides a principled approach to quantifying uncertainty in model parameters and model structure and has seen a surge of applications in recent years. 
Within the context of a Bayesian workflow, we are concerned with model selection for the purpose of finding models that best explain the data, that is, help us understand the underlying data generating process. Since we rarely have access to the true process, all we are left with during real-world analyses is incomplete causal knowledge from sources outside of the current data and model predictions of said data. This leads to the important question of when the use of prediction as a proxy for explanation for the purpose of model selection is valid.
We approach this question by means of large-scale simulations of Bayesian generalized linear models where we investigate various causal and statistical misspecifications. Our results indicate that the use of prediction as proxy for explanation is valid and safe only when the models under consideration are sufficiently consistent with the underlying causal structure of the true data generating process.
\end{abstract}


\input{content}

\section*{Funding}
This work was partially funded by the Deutsche Forschungsgemeinschaft (DFG, German Research Foundation) under Germany’s Excellence Strategy -- EXC-2075 - 390740016 (the Stuttgart Cluster of Excellence SimTech).

This work was performed on the computational resource bwUniCluster funded by the Ministry of Science, Research and the Arts Baden-Württemberg and the Universities of the State of Baden-Württemberg, Germany, within the framework program bwHPC.

The authors gratefully acknowledge the support and funding.

\section*{Acknowledgements}
We also want to thank Marvin Schmitt and Stefan Radev for their feedback on earlier versions of this manuscript as well as Yannick Dzubba for his help during development of the software used for our simulation study.

\section*{Data availability}
The data that support the findings of this study are openly available in OSF at \url{http://doi.org/10.17605/OSF.IO/XGKZV}.

\bibliographystyle{unsrtnat}
\bibliography{references}  

\end{document}

%% file: content.tex
\section{Introduction}
\label{sec:introduction}

Probabilistic modeling provides a principled approach to quantifying uncertainty in model parameters and model structure. 
In recent years, it has seen a surge of applications in almost all quantitative sciences and industrial areas \citep{gelman_bayesian_2013, mcelreath_statistical_2020, gelman_bayesian_2020}. 
To support the principled application of Bayesian methods, \cite{gelman_bayesian_2020} proposed an overarching workflow to conduct Bayesian data analysis.
In short, the workflow asks users to pick an initial model and iteratively refine it, performing various checks on the way to ensure that probabilistic assumptions are sensible, computations are valid, and model results are trustworthy for the intended purposes.
This basic model building loop is repeated until either the user's requirements are satisfied or no satisfactory model can be found with the available resources.
Within this overarching workflow, there are still many unknowns when it comes to making optimal decisions for each iterative step or sub-workflow.

The current work is concerned with the decision-making process during model building iterations, namely model selection for the purpose of finding models that best \textit{explain} the data, that is, help us understand the underlying data generating process. Unfortunately, explanation is an illusive concept in practice as, for real data, we have no direct access to the underlying data generating process. Thus, all we are left with during real-world analyses is incomplete \textit{causal} knowledge from sources outside of the current data and model \textit{predictions} of said data. The distinction between explanation (of the underlying process) and prediction (of observable data) has a long history in many fields of sciences engaged in model-based reasoning and inference (see Section \ref{sec:explanation-prediction}). However, the statistical relationship of explanation and prediction is still not sufficiently understood, which is surprising given the importance of these concepts to statistical theory and practice. 

\begin{figure}[tb]
\centering
\includegraphics[width=1\linewidth]{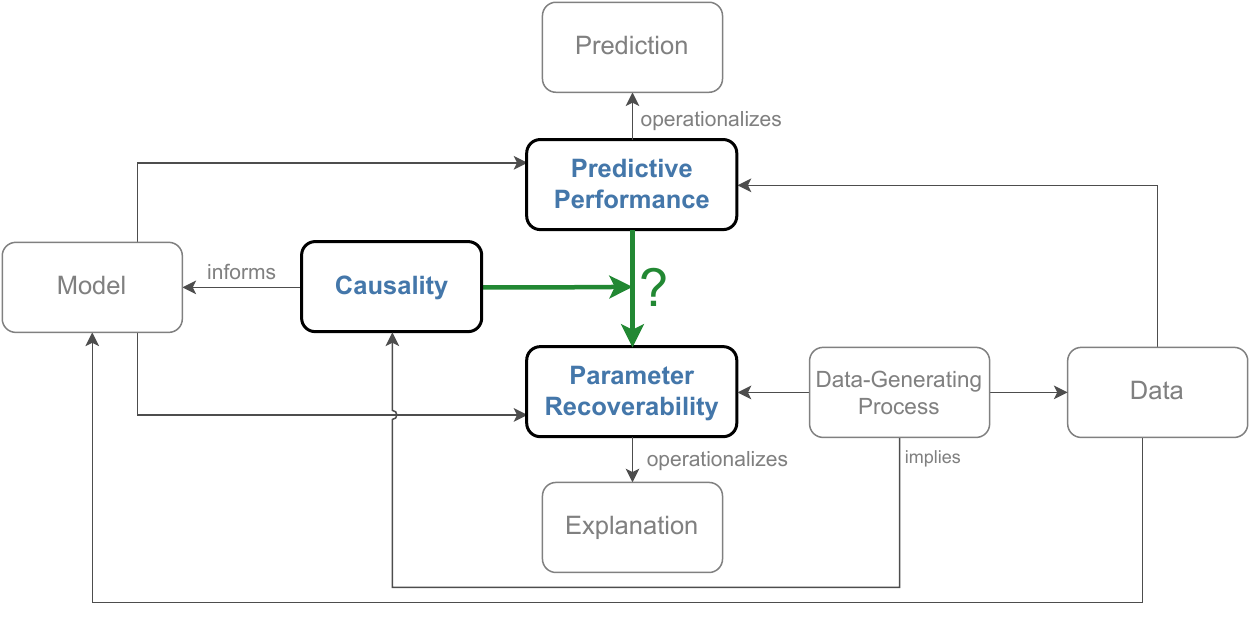}
\caption{High-level overview of our conceptual framework that connects explanation, prediction and causality in statistical modeling. Parameter recovery compares model parameters and the data-generating process to operationalize explanation.
Predictive performance compares model predictions and observed data to operationalize prediction. Causality is implied by the data-generating process and can inform model building. The three-way relationship between predictive performance, parameter recovery, and causality is unclear and subject of this paper. See Section~\ref{sec:explanation-prediction} for details.}
\label{fig:intro-figure}
\end{figure}

To this end, we study the validity of using prediction as a proxy for explanation in Bayesian statistical models (see Figure~\ref{fig:intro-figure} for a conceptual overview to be detailed in Section~\ref{sec:explanation-prediction}). Our main conclusion can be summarized as follows: 
\textit{Using prediction as a proxy for explanation is valid and safe only when the considered models are sufficiently consistent with the underlying causal structure of the true data generating process}. Specifically, our contributions are:

\begin{enumerate}[(i)]
    \item a conceptual introduction and overview of the relationship of explanation and prediction as well as their connection to causality;
    \item large-scale simulations of Bayesian generalized-linear models to study said relationship under various causal and statistical misspecifications;
    \item initial evidence that causality is indeed the missing link that connects prediction and explanation when comparing statistical models; and
    \item a set of R packages \citep{scholz_bayesim_2022, scholz_bayesfam, scholz_bayeshear} that facilitate simulation studies of Bayesian models.
\end{enumerate}

\subsection{Explanation vs. Prediction}
\label{sec:explanation-prediction}

\cite{douglas2009reintroducing} provides a historical overview of the relationship of prediction and explanation from the perspective of philosophy of science. She concludes that 'explanations are the means that help us think our way through to the next testable prediction' \citep[page 462, Section 6]{douglas2009reintroducing}. In other words, and more statistically speaking, explanations help us understand the inner workings of a process by means of a model, while prediction concerns the comparison of the model's output to real-world observables.
Explanation and prediction are not statistical concepts themselves, though. Instead, we have to operationalize them in terms of mathematical quantities computable from a statistical model. Following \cite{buerkner_utility_2022}, we call the statistical operationalization of explanation \textit{parameter recoverability} and the operationalization of prediction \textit{predictive performance}:

Parameter Recoverability (PR) refers to a model's ability to recover the latent (i.e., not directly observable) parameters of an assumed true data-generating process (DGP), for example, the true effect size of a treatment for a medical condition \citep{buerkner_utility_2022}. 
This implies that assessing PR requires concrete knowledge of the true DGP including its true parameter values.
Accordingly, it can only be studied directly in artificial settings where the true DGP is known, with the hope that the real data of interest fulfills the assumptions of these artificial settings sufficiently well.

Predictive Performance (PP) describes the ability of a model to accurately predict new observations from existing observations, for example, predicting a diagnosis based on symptoms and medical history \citep{hastie_elements_2009, gabaix2008seven}.
It is one of the most prominent utilities for gauging model performance in data analysis and a central tool for model comparisons \citep{vehtari_survey_2012, gelman_bayesian_2013, vehtari_practical_2017}.
In most cases, one is interested in out-of-sample PP, as predicting unseen data is almost always the primary goal of modeling \citep{vehtari_survey_2012}.
PP is conceptually similar to PR in that both target the accurate estimation of model-implied quantities \citep{buerkner_utility_2022}, with the main difference that the former targets quantities that are observable at real data inference time (i.e., outcome variables), which allows to derive estimates that are agnostic with respect to the true DGP \citep{vehtari_survey_2012}.

In the following, we present some of the common perspectives on prediction and explanation as well as their connections with causality.
When talking about these concepts from statistical perspectives, we use them and their operationalizations interchangeably.

\subsubsection{Explanation only}
\label{sec:explanation}

The statistical field that is probably most interested in explanation in the form of unbiased estimators is that of causality \citep{pearl2009causality}.
The \emph{do}-calculus \citep{pearl_do-calculus_2012} and equivalently the potential outcomes framework \citep{rubin1974estimating, rubin1978bayesian} offer sound theories to identify models that are able to provide an unbiased estimate of a parameter of choice.
We call a model 'causally consistent' if it is consistent with a theoretically justified causal graph of its contributing variables \citep{buerkner_utility_2022}.
This consistency is necessary for models to provide at least asymptotically unbiased estimates for latent parameters. 

While the ability to find an unbiased estimator is highly valuable, it is not the only metric we need to pay attention to.
As argued for example by \cite{shmueli_explain_2010}, optimizing the overall bias-variance trade-off is a sensible alternative goal, which also requires to minimize the variance of a parameter estimator in addition to minimizing its bias.
Examples in the causal literature that attempt to additionally reduce variance are the works of \cite{henckel2022graphical} and \cite{rotnitzky2020efficient}, which use graphical criteria to find lower variance estimators \citep[see also][for a more general overview]{cinelli_crash_2020}.
However, we found that the reduction of variance is only considered after unbiasedness has been established. Thus, no actual bias-variance trade-off is allowed to take place from this perspective.
Additionally, predictive performance considerations are missing altogether from the general discussion in the causal literature.
We go into more details about how causal consistency serves as the foundation of the presented study in Section \ref{sec:causal_foundation}.

\subsubsection{Explanation first}
\label{sec:explanation-first}

Arguably the most prominent perspective on the relationship of explanation and prediction is coming from the philosophy of science \cite[as examplarily presented by][]{ramspek2021prediction, navarro_between_2019, hernan2019second, yarkoni_choosing_2017, shmueli_explain_2010, breiman_statistical_2001} in cognitive-, social-, and life-sciences.
The main discussion point is that, even if explanation is the main research goal, the value of good prediction shouldn't simply be ignored.
\cite{yarkoni_choosing_2017} and \cite{breiman_statistical_2001} are examples for the call to put more focus on predictions, even if explanation is the main goal.
The underlying argument is that a model with bad predictive performance has a higher chance of misrepresenting or missing relevant features of the true DGP. One should thus be careful when relying on the explanations of a model with weak predictive performance.
Similarly, \cite{shmueli_explain_2010} makes the argument that predictive power cannot be inferred from explanatory power, that is, good explanatory models do not automatically lead to good predictions.
A common proposal is to use a predictive model (i.e., a model that maximizes predictive performance) as a benchmark \citep{shmueli_explain_2010, cranmer2017can} to aim for with an explanatory model (i.e., a model that maximizes parameter recoverability).
\cite{douglas2009reintroducing} gives an illustrative example for this approach with the discovery of Pluto. The paths of Uranus and Neptune did not follow their predicted paths which implied a problem with the explanatory model. The solution was the discovery of Pluto, which explained the observed deviation and lead to an updated model.
Finally, while there are sound high level arguments for why models that provide better explanations should generally provide better predictions as well, we are not aware of other works that investigate this relationship further besides common textbook examples of improved predictions despite bad parameter recovery in the face of causal misspecification \citep{mcelreath_statistical_2020}. 

\subsubsection{Prediction first}
\label{sec:prediction-first}

Machine-learning models are sometimes also called black box models due to the fact that they solely focus on providing the best possible predictions while their inner workings are hard or impossible to understand and do not directly map to any assumed DGP.
The lack of interpretability, and thus potential to provide explanations, is a common critique of machine learning techniques, more recently more work has been done to improve this aspect \citep[see for example][]{gilpin2018explaining, carvalho2019machine}.
While predictive performance remains the main goal of these models, requirements regarding responsible reporting and accountability as well as improved performance for small sample sizes led to the incorporation of causal assumptions, which are a requirement for reliable explanations \citep{li2020incorporating, richens2020improving, scholkopf2022causality}.
And while not necessarily using the same causal vocabulary, the idea of physics-informed neural networks \citep{raissi2019physics, karniadakis_physics-informed_2021} is closely related, as the partial or ordinary differential equations describing physical processes can be understood as causal constraints that have to be met.

However, with the increasing requirements for explainable AI \citep{arrieta2020explainable} in fields like medicine, aspects such as feature importance \citep{lundberg2017Unified, lime} become more relevant to understand how machine learning models make decisions. Such aspects can be seen as extending the concept of explanation in some sense, although this has not been explored systematically to our knowledge and, at least in the stricter sense of recovering (latent) parameters of an assumed DGP, explanation is not really a concern in machine learning in general.

\subsubsection{Combining explanation, prediction, and causality}
\label{sec:combining-prediction-explanation}

Based on the presented perspectives, there is an apparent gap in the joint consideration of explanation, prediction, and causality. We illustrate this gap in Figure~\ref{fig:intro-figure}, where we provide an overview of these concepts' relationships in a statistical modeling workflow.
In this paper, we approach their joint consideration specifically from the perspective of Bayesian model selection. 

When explanation is the goal, prediction reduces to a conveniently available supporting utility that \emph{ideally} helps to select better explaining models at real data inference time \citep{buerkner_utility_2022}. 
In practice at least, and despite the theoretical arguments for caution, this assumption is very commonly (and often implicitly) made whenever explanatory model choices are based on out-of-sample posterior predictive metrics or their approximations, such as AIC \citep[e.g.,][]{watanabe_algebraic_2009, mcelreath_statistical_2020}, DIC \citep{spiegelhalter_bayesian_1998}, WAIC \citep{watanabe_asymptotic_2010, vehtari_practical_2017} or ELPD-LOO \citep{vehtari_survey_2012, vehtari_practical_2017}.
As we know from counterexamples \citep{betancourt_towards_2020, hagerty1991comparing}, this assumption cannot hold in general, but it remains unclear under which conditions it is actually justified.
The goal of this paper is to investigate under which circumstances this proxy use might be valid.

\subsection{Generalized Linear Models}
\label{sec:glm}

We focus our studies on the class of Bayesian generalized linear models (GLMs). This choice is motivated by the fact that they represent the minimal general class of models that enable us to investigate the utility of using prediction as a proxy for explanation.
GLMs allow us to represent causal DAGs while adding the ability to change model aspects that are not DAG-dependent (i.e., the likelihood family and link function).
Despite (or perhaps because of) their simplicity, GLMs make up a big part of all statistical data analyses. Their success can be explained by several factors, involving the ease of interpretation of their additive structure, rich mathematical theory, and (relatively) straightforward estimation \citep{gelman_regression_2020, harrell_regression_2015, nelder_generalized_1972, gill_generalized_2001}.

More specifically, we consider GLMs of the form $y \sim \textrm{likelihood}(\textrm{link}^{-1}(\mu),\phi)$ with a univariate response variable \(y\) that is assumed to follow a parametric likelihood distribution, often called a likelihood family \citep{bates_fitting_2015, burkner2017brms}, one main location parameter $\mu = \alpha + X\beta$ that is predicted, as well as zero or more auxiliary distributional parameters $\phi$ that are assumed to be constant over observations \citep{nelder_generalized_1972, gill_generalized_2001}.

When setting up GLMs, the four important choices the analyst has to make are (i) the likelihood family, (ii) the link function, (iii) the linear predictor term, and (iv) whether and how to regularize, that is, for Bayesian GLMs, what prior distributions to use. All of these choices are mutually related \citep{gelman_prior_2017}, but specifically (i) and (ii) are closely intertwined, as the choice of link function depends on the support of $\mu$ and thus on the chosen likelihood. 
We discuss our specific choices of likelihoods and link functions in Section \ref{sec:likelihoods_and_links}, the specific models we generate data from in Section \ref{sec:data-generation} and the models we fit on the generated data in Section \ref{sec:model-fitting}.

\subsection{Aims and Scope}
\label{sec:aim_and_structure}

As laid out in Section \ref{sec:explanation-prediction}, there is a gap in the literature regarding the relationship of (the statistical operationalizations of) prediction and explanation under varying causal assumptions.
The aim of this paper is not to fill said gap entirely but rather to be a starting point for further research.
As explained above, direct measures of explanation are rarely accessible during real-world modeling and prediction provides a convenient (i.e., easily available) alternative for model selection.
Even though the validity of the practice has so far remained unclear, prediction is commonly used as a proxy for explanation during practical model selection in the context of Bayesian or other statistical workflows.
It is the main aim of this paper to empirically investigate the use of prediction as a proxy for explanation under several causal and statistical misspecification mechanisms in Bayesian GLMs.
Further, the methods introduced in this paper could serve as a basis for further work, some of which we propose in Section \ref{sec:discussion}.

While this paper focuses on Bayesian models, we assume that the presented results also hold for their frequentist counterparts, as the mechanisms of causality are agnostic to the specific method of estimation (under reasonable equivalence of estimators; see also Section \ref{sec:model-fitting}).

\section{Methods}
\label{sec:methods}

As explained in Section \ref{sec:explanation-prediction}, studying parameter recoverability (PR; the statistical operationalization of explanation) requires knowledge of the true data-generating process (DGP). Combined with the fact that we aim to study Bayesian GLMs, this renders an analytical approach infeasible and we thus resort to extensive simulations as the method of choice.

\begin{figure}[tb]
\centering
\includegraphics[width=1\linewidth]{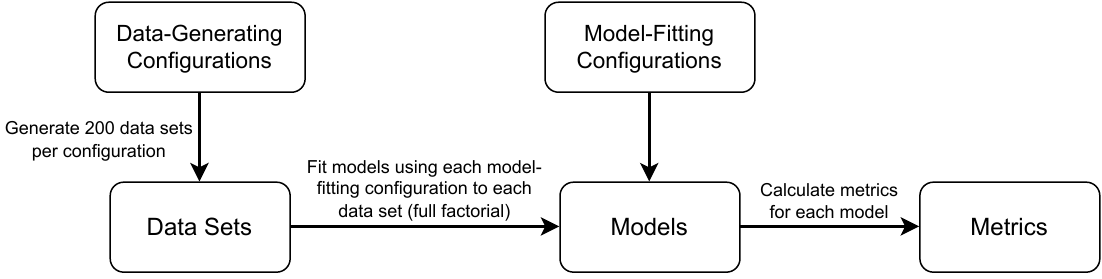}
\caption{Conceptual simulation architecture. We generated 200 data sets for each entry from a list of data-generating configurations (see Table~\ref{tab:data-gen-conf}) for a total of $28,800$ datasets. We then fit a model for each entry from a list of model-fitting configurations (see Table~\ref{tab:model-fit-conf}) to each data set for a total of $1,728,000$ models. Finally, we calculated metrics for all fitted models.}
\label{fig:simulation}
\end{figure}

In this section, we start with an explanation of the causal foundation of the simulations and how we used it throughout the process in Section \ref{sec:causal_foundation}.
We also provide an overview of the considered likelihood families and link functions in Section \ref{sec:likelihoods_and_links}.
Next, we walk through the study's design (see Figure~\ref{fig:simulation} for a conceptual overview), with data generation described in Section \ref{sec:data-generation}, model fitting in Section \ref{sec:model-fitting}, and metric calculation in Section \ref{sec:model-based-metrics}.
Finally, we discuss the statistical analysis of the results of the simulations in Section \ref{sec:statistical-analysis}.

Many aspects of this study are not only dependent on the methods and algorithms used, but also on their software implementations.
The simulation was implemented in R \citep{team2013r} using Stan \citep{stan_2022, carpenter2017stan} and brms \citep{burkner2017brms}. Our software packages bayesim \citep{scholz_bayesim_2022}, bayesfam \citep{scholz_bayesfam}, and bayeshear \citep{scholz_bayeshear} are available online.
A more thorough discussion of the included likelihoods and links as well as additional results are available in the supplemental material.
Said discussion as well as the code for the simulation configurations, the simulation data and corresponding analyses are available in our online appendix \citep{online_appendix}.

\subsection{Causal Foundation}
\label{sec:causal_foundation}

\begin{figure}[tb]
\centering
\includegraphics[width=\linewidth]{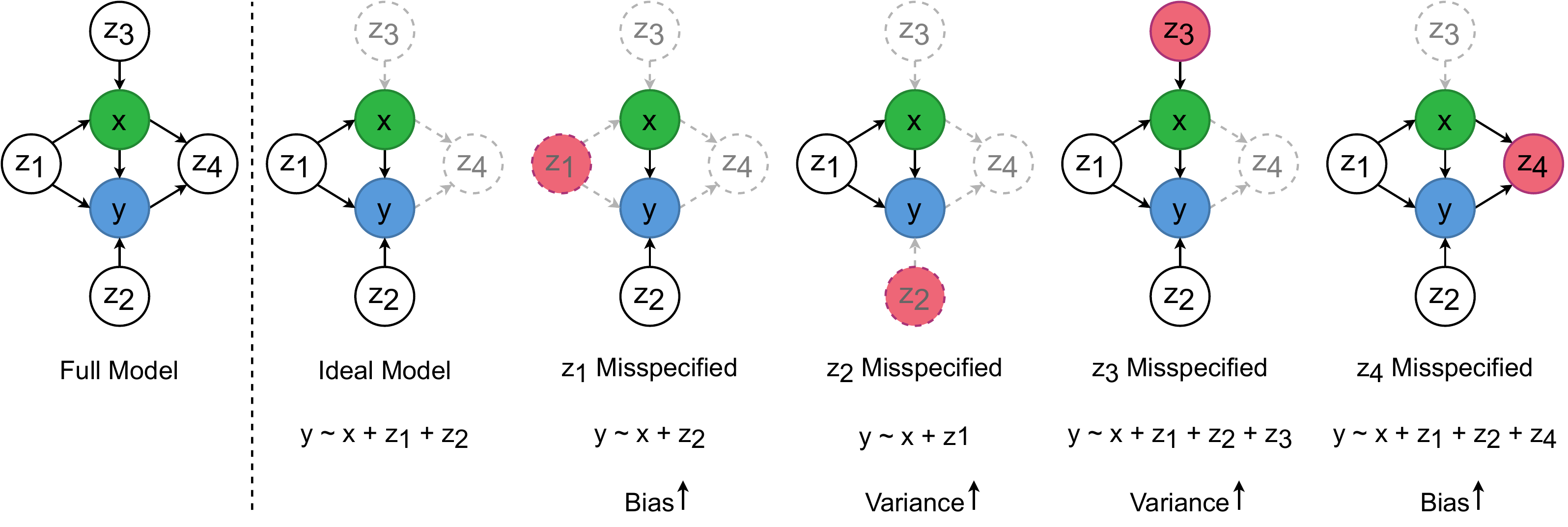}
\caption{Full data-generating DAG and the resulting models. The ideal model is the subset of the full model that optimally estimates $\beta_{xy}$. Misspecifying with respect to each of the $z_n$ variables leads to an additional model, where excluding $z_1$ or including $z_4$ increases bias, while excluding $z_2$ or including $z_3$ increases variance of the estimation.} 
\label{fig:dag}
\end{figure}

In practice, we usually cannot know if a model actually represents the true DGP,
not only in terms of the distributional assumptions of likelihood and link, but also in terms of the structural and causal assumptions of the included (causal) covariates and their (non-)linear combination on the latent space.
This uncertainty about covariates and their causal effects is a focus point for the causal literature. As explained in Section \ref{sec:explanation}, causal consistency is a necessity for good parameter recoverability and thus a fundamental aspect of this study's methods.

Here, we use a single, prototypical causal directed acyclic graph \citep[DAG;][]{pearl2009causality} to guide both data generation and causal model assumptions (see Figure~\ref{fig:dag}).
The full DAG displayed on the left-hand side of Figure~\ref{fig:dag} consists of an outcome $y$, a treatment $x$, and four additional variables $z_1, z_2, z_3, z_4$ that correspond to four common, qualitatively different types of controls \citep{cinelli_crash_2020}. With respect to the effect of $x$ on $y$, $z_1$ is a fork, $z_2$ is an ancestor of $y$, $z_3$ is an ancestor of $x$, and $z_4$ is a collider. These archetypes represent the majority of controls occurring in reality\citep{cinelli_crash_2020, pearl2009causality}. The only common control missing from this list is the pipe $(x \rightarrow z \rightarrow y)$, which was not included here due to its similarity with the fork.

The goal of our statistical models will be to estimate the causal effect $\beta_{xy}$  of $x$ on $y$ as accurately as possible. 
For this purpose, $z_1$ constitutes a 'good control' in the sense that controlling for it decreases bias in the estimation, while $z_4$ is a 'bad control', as controlling for it increases bias.
In contrast, $z_2$ and $z_3$ are 'neutral controls': controlling for them does not influence bias. However, controlling for $z_2$ increases precision (decreases variance) while controlling for $z_3$ decreases precision of the estimation of $\beta_{xy}$. As a result, to estimate  $\beta_{xy}$, the \textit{ideal (causal) model} only controls for $z_1$ and $z_2$ (see the second graph in Figure~\ref{fig:dag}). From there, we can obtain \textit{misspecified (causal) models} by including or excluding controls. Based on four controls, $2^4 - 1 = 15$ misspecified models are conceivable, but for simplicity, we focus only on a subset of four of them, each of which deviates from the ideal model in exactly one control (see the third to sixth graph in Figure~\ref{fig:dag}). Two of these misspecified models (i.e., when incorrectly excluding $z_2$ or incorrectly including $z_3$) still yield (asymptotically) unbiased estimates of $\beta_{xy}$. We will call them and the ideal causal model 'causally unbiased', while we call (asymptotically) biased models 'causally biased' \citep{buerkner_utility_2022}.
Using this terminology, we can formulate our primary research questions more precisely as: 
\begin{enumerate}[(i)]
    \item\label{enum:research-questions:PPPR} Within a set of statistical models that all share the same underlying causal model, can prediction be reliably used as a proxy for explanation?
    \item Does the answer to (\ref{enum:research-questions:PPPR}) depend on whether or not the causal model is biased?
\end{enumerate}

Focusing on these questions implies that we do not aim to compare models with differing linear predictor terms (i.e., causal assumptions as per Figure~\ref{fig:dag}).
For one, there already are well known examples of cross-formula comparisons where the alignment of prediction and explanation doesn't hold \citep[e.g., adding a collider improves prediction but worsens explanation compared to an ideal unbiased model;][]{mcelreath_statistical_2020}. Still, cross-formula comparison includes a lot of uncharted territory, which we think is worth studying but out of scope of the present paper.
Of course, as a result of this choice, we need other aspects to vary among the compared statistical models. In this study, these aspects will be the likelihood and link functions of Bayesian GLMs as detailed next.

\subsection{Likelihoods and Link Functions}
\label{sec:likelihoods_and_links}

The range of practically relevant likelihood classes is extensive and encompasses, among others, likelihoods for unbounded, lower-bounded, and double-bounded continuous data, as well as binary, categorical, ordinal, count, and proportional (sum-to-one) data \citep{johnson_continuous_1995, johnson_discrete_2005, stasinopoulos_gamlss_2007, yee_vgam_2010, buerkner_bayesian_2021}. Studying all of them at once would be too large of a scope for a single paper. Here, we focus our efforts on GLMs with lower-bounded or double-bounded continuous likelihoods. Within these classes, we not only have several qualitatively different (non-nested) likelihood options, but can also study both main classes of non-identity link functions.

Below, we give a short overview of the likelihood and link functions included in the simulations.
For consistency, we use mean parameterizations of likelihoods throughout or, if those are unavailable, median parameterizations. 
A more detailed review of the considered options, our inclusion criteria, and the used parameterizations are available in the supplemental material and online appendix \citep{online_appendix}.

\subsubsection{Likelihoods and links for double-bounded responses}
\begin{figure}
\centering
\includegraphics[width=0.99\linewidth]{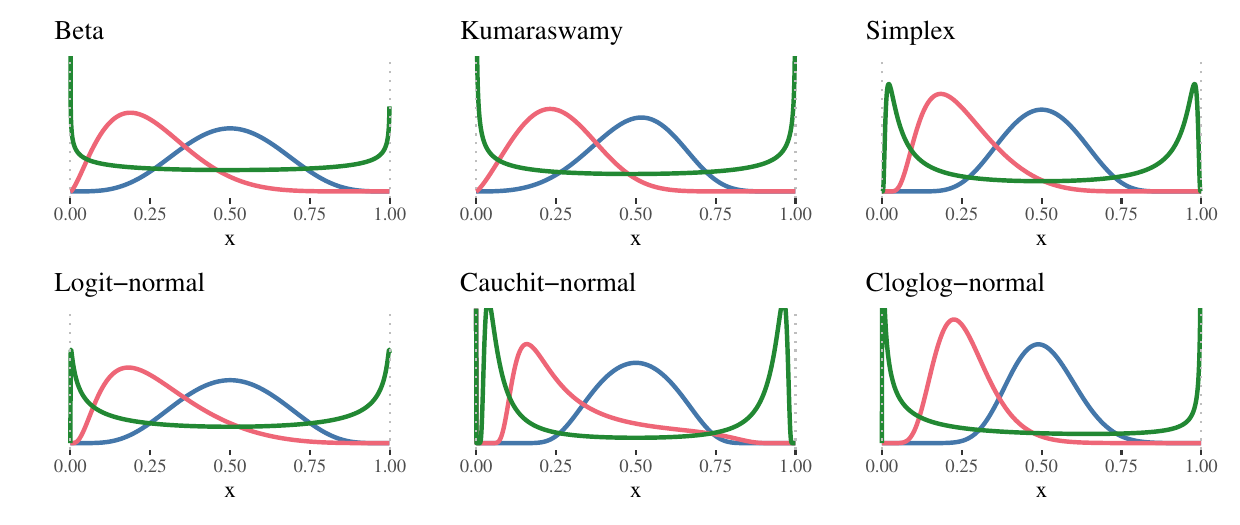}
\caption{Example illustrations of all included double-bounded densities each with three different shapes. The shapes result from different distributional parameters (detailed in the online appendix \cite{online_appendix}). The parameters were specifically chosen to produce the three qualitatively different shapes accommodated by the different likelihoods, namely a symmetric, an asymmetric and a bimodal bathtub shape. The y-axis is truncated at 5 from above for better visibility of the different shapes.} 
\label{fig:unit_interval_densities}
\end{figure}

Without loss of generality, any double-bounded response can be linearly transformed to the unit interval. Accordingly, it is sufficient to focus on likelihoods for unit interval data.
We included the beta \citep{espinheira_beta_2008}, Kumaraswamy \citep{kumaraswamy_generalized_1980}, simplex \citep{barndorff-nielsen_parametric_1991}, and transformed-normal \citep{atchison_logistic-normal_1980, kim_sample_2017} likelihoods.
The transformed-normal likelihoods arise from applying the link functions to the response variable $y$ (e.g., resulting in the logit-normal likelihood), instead of to the location parameter $\mu$ as would be usual in standard GLMs. All of these likelihoods have two distributional parameters, one location (mean or median) and one scale/shape parameter, only the former being predicted as per the causal graph structures.
Figure~\ref{fig:unit_interval_densities} shows some example densities for each likelihood, illustrating qualitatively different shapes they can accommodate.
The three distinct shapes are unimodal symmetric and asymmetric shapes as well as a bimodal bathtub shape.
For the remainder of the paper, we will refer to these shapes as symmetric, asymmetric, and bathtub, respectively.
As link functions, we included the logit, cloglog, and cauchit links, each of them having qualitatively different properties \citep{yin_skewed_2020, jiang_new_2013, lemonte_new_2018, damisa_comparison_2017, fahrmeir_multivariate_1994, gill_generalized_2001, powers2008statistical, morgan_note_1992, koenker_parametric_2009, lemonte_new_2018}.
The logit link is based on the symmetric, light-tailed logistic distribution, the cloglog link is based on the asymmetric Gumbel distribution, while the cauchit link is based on the symmetric, heavy tailed Cauchy distribution.
Since logit and probit yield almost indistinguishable results due to the similar shapes of the logistic and normal distributions \citep{fahrmeir_multivariate_1994, gill_generalized_2001, powers2008statistical}, we decided against including the probit link despite its prominence.

\subsubsection{Likelihoods and links for lower-bounded responses.}
\begin{figure}[tb] 
\centering
\includegraphics[width=0.99\linewidth]{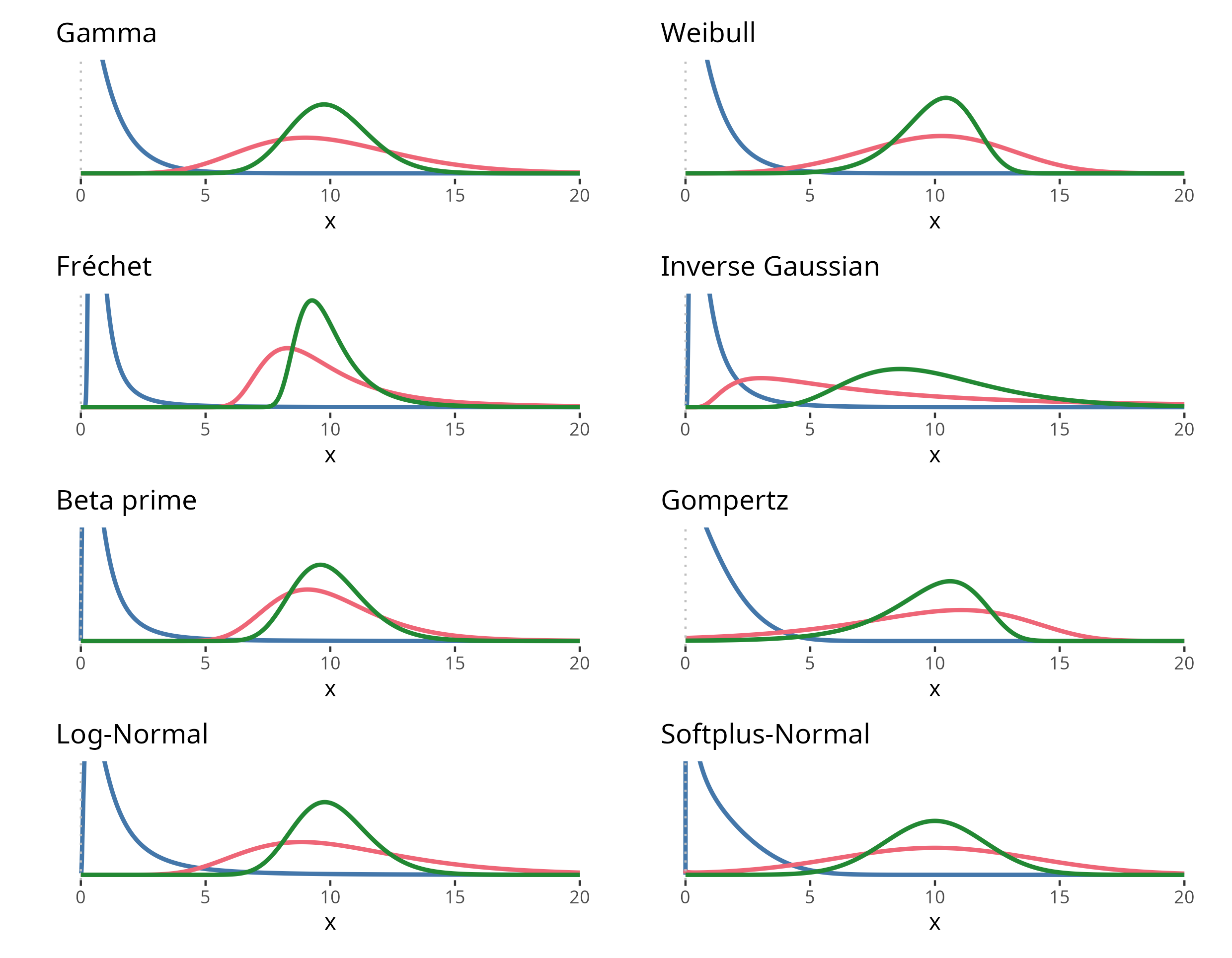}
\caption{Example illustrations of all included lower-bounded densities, each with three different shapes. The shapes result from different distributional parameters (detailed in the online appendix \cite{online_appendix}). The parameters were specifically chosen to produce the three qualitatively different shapes accommodated by the different likelihoods, namely a more symmetric, a more asymmetric and a ramp shape. The y-axis is truncated at 0.4 from above for better visibility of the different shapes.} 
\label{fig:lower_bounded_densities}
\end{figure}

Without loss of generality, any continuous lower-bounded response can be linearly transformed to have a lower bound of zero. Accordingly, it is sufficient to focus on likelihoods for strictly positive data.
We included the gamma, Weibull, Fréchet, inverse Gaussian, beta prime, Gompertz, and transformed-normal likelihoods\footnote{In preliminary analysis, we observed consistently bad posterior sampling behavior of inverse-Gaussian models, with very slow sampling overall and more than half of the models failing to converge.
For these reasons we decided to exclude the inverse-Gaussian likelihood from the final analysis.}
all of whom have two distributional parameters, namely location (mean or median) and scale or shape).
Figure~\ref{fig:lower_bounded_densities} shows example densities for each likelihood, illustrating qualitatively different kinds of shapes they can accommodate.
The three distinct shapes are unimodal thin tail and heavy tail shapes as well as a ramp shape. For the remainder of the paper we will refer to these shapes as thin tail, heavy tail, and ramp, respectively.
As link functions, we included the log and the softplus link. In contrast to the multiplicative log link,
softplus approaches the identity for larger values, thus approximating additive behavior of regression terms while enforcing positive predictions at the same time \citep{zheng2015improving, dugas2000incorporating}.

\subsection{Data Generation}
\label{sec:data-generation}

For each of the combinations of likelihood family, likelihood shape, and link presented in Section \ref{sec:likelihoods_and_links}, we simulated data sets based on the full DAG (Figure~\ref{fig:dag} left-hand side) as follows:
\begin{align*}
z_1 & \sim \mathrm{normal}(0, \sigma_{z_1}), \quad
z_2 \sim \mathrm{normal}(0, \sigma_{z_2}), \quad
z_3 \sim \mathrm{normal}(0, \sigma_{z_3}) \\
x & \sim \mathrm{normal}(\beta_{z_1x}z_1 + \beta_{z_3x}z_3, \sigma_x) \\
y & \sim \textrm{likelihood}(\textrm{link}^{-1}(\alpha_y + \beta_{xy}x + \beta_{z_1y}z_1 + \beta_{z_2y}z_2), \phi) \\
z_4 & \sim \mathrm{normal}(\beta_{xz_4}x + \beta_{yz_4}y, \sigma_{z_4}).
\end{align*}
Here, $\phi$ denotes the second distributional (scale or shape) parameter of the specified likelihood family and $\textrm{link}^{-1}$ denotes the inverse link (or response-) function. We chose normal distributions to generate all data variables besides $y$ to limit the scope of our simulations. Each simulated data set contained $100$ observations.
Since the fitted models only have between 4 and 6 parameters, they are simple enough to be well identified on the basis of $100$ observations alone.
The parameters of the true DGP were set to fixed values, rather than being drawn from a prior distribution \citep{talts_validating_2020}, to avoid blurring the effect of the induced causal misspecifications and different likelihood shapes. This decision was made due to the extreme data sets that prior distributions over parameters could create, which is a general challenge for Bayesian simulations \citep{gabry_visualization_2019, mikkola_prior_2021}.

Each likelihood's second distributional parameter $\phi$ as well as the intercept $\alpha_y$ were chosen to produce the shapes presented in Section~\ref{sec:likelihoods_and_links}. 
The individual coefficients for $x$ and the $z_i$ were calibrated so that the parameter recovery was imperfect for the ideal model while also preventing the causally misspecified models from consistently failing (see also Section \ref{sec:model-fitting}). The true causal effect $\beta_{xy}$ of $x$ on $y$ was either fixed to zero or set to a non-zero value that was calibrated together with all other coefficients.
To prevent response values from under- or overflowing to the lower- or upper boundaries numerically, we truncated them near the boundaries around the 6$^\text{th}$ digit.

For each data generation configuration implied by fully crossing the design factors (see Table \ref{tab:data-gen-conf}), we generated $200$ data sets, which resulted in $14,400$ data sets each for the double- and single-bounded scenarios.

\begin{table}[tb]
\caption{Overview of data generation configurations.}
  \centering
    \begin{tabularx}{\linewidth}{ l X }
    \hline 
    Factor & Levels \\
    \hline 
    Double-bounded likelihoods & beta, Kumaraswamy, simplex, transformed-normal \\ 
    Double-bounded links & logit, cauchit, cloglog \\ 
    Double-bounded shapes & symmetric, asymmetric, bathtub \\ 
    Lower-bounded likelihoods & gamma, Weibull, transformed-normal, Fréchet, beta-prime, Gompertz \\ 
    Lower-bounded links & log, softplus \\ 
    Lower-bounded shapes & ramp, heavy tail, thin tail \\
    True $\beta_{xy}$ & zero, positive \\ \hline
    \end{tabularx}%
  \label{tab:data-gen-conf}%
\end{table}%
                                
\subsection{Model Fitting}                              
\label{sec:model-fitting}

On each generated data set, we fitted all models resulting from the fully crossed combination of likelihoods and links (see Section \ref{sec:likelihoods_and_links}) as well as the five different linear predictor terms implied by the ideal and misspecified causal models (see Section \ref{sec:causal_foundation}). In reference to R formula syntax, we will also refer to the different linear predictor terms as \emph{formulas} in the following.
Accordingly, given a data set generated from a double-bounded likelihood, we fitted models on that data set using all combinations of double-bounded likelihoods, links, and DAG-based formulas (see Figure~\ref{fig:dag}). The same approach was followed for the lower-bounded data. An overview of the model fit configurations is given in Table \ref{tab:model-fit-conf}.

The fully-crossed design results in $60$ fit configurations for the double-bounded and lower-bounded models.
Multiplied with $14,400$ data sets each, this leads to a total of $864,000$ double-bounded and single-bounded models each fitted in our simulations.

\begin{table}[tb]
  \centering
  \caption{Overview of model fit configurations.}
    \begin{tabularx}{\linewidth}{ l X }
    \hline
    Factor & Levels \\
    \hline 
    Double-bounded Likelihoods & beta, Kumaraswamy, simplex, transformed-normal \\ 
    Double-bounded Links & logit, cauchit, cloglog \\ 
    Single-bounded Likelihoods & gamma, Weibull, transformed-normal, Fréchet, beta-prime, Gompertz \\ 
    Single-bounded Links & log, softplus \\ 
    Formulas (right-hand side) & $x + z_1 + z_2$, $\; x + z_2$, $\; x + z_1$, $\; x + z_1 + z_2 + z_3$, $\; x + z_1 + z_2 + z_4$\\ \hline
    \end{tabularx}%
  \label{tab:model-fit-conf}%
\end{table}%

Contrary to what we would recommend in practical applications of Bayesian models, we used flat priors for all model parameters, as it is not clear to us how one would specify equivalent priors for the different auxiliary parameters $\phi$ across all likelihood.
Additionally, different links imply different latent scales, which render the regression coefficients' scales incomparable across (assumed) links and thus further complicates equivalent prior specification. 
In a real-world analysis, we would prefer to use at least weakly-informative priors \citep{stan_2022, gelman_bayesian_2013, mcelreath_statistical_2020}.
Accordingly, the current choice is to be understood only in the context of the present simulation study aiming to ensure comparability across likelihoods and links.
In pilot experiments (not shown here), we have confirmed that the differences in posteriors as well as the implied prediction metrics between models with flat vs. weakly informative priors are minimal for the models under investigation.
We argue that such minimal differences do not justify extensive evaluation of different prior choices, since this is not the focus of the present paper.
Additionally, the use of flat priors results in model estimations very similar to maximum likelihood estimation, which is another reason why we expect the results of this study to generalize to frequentist models as well.  

All models were fitted using Stan \citep{carpenter2017stan, stan_2022} via brms \citep{burkner2017brms} with two chains, 500 warmup- and 2000 post-warmup samples, which resulted in $4000$ total post-warmup posterior samples per model.
We used an initialization range of $0.1$ around the origin on the unconstrained parameter space to avoid occasional initialization failures.
For all other MCMC hyperparameters, we applied the brms defaults \citep{burkner2017brms}.

\subsection{Model-Based Metrics}
\label{sec:model-based-metrics}

To measure parameter recovery and predictive performance of each fitted model, we used multiple metrics as detailed below. Implementations of these metrics are provided in the R packages loo \citep{Vehtari2022looR}, posterior \citep{burkner2022posteriorR}, bayesim \citep{scholz_bayesim_2022}, and bayeshear \citep{scholz_bayeshear}.

F\subsubsection{Parameter recoverability}

To date, the primary metric for causal parameter recoverability (operationalizing explanation) remains the estimation bias.
However, as argued in Section \ref{sec:explanation-prediction}, it may not be the only sensible metric. Instead, in order to allow for a bias-variance trade-off in the causal estimate's evaluation, the RMSE may be a worthy alternative metric in non-asymptotic regimes.

More precisely, given a true parameter value $\theta$ and a set of corresponding posterior samples $\{ \theta^{(s)} \}$, we compute the sampling-based posterior bias and $\mathrm{RMSE}$ as
\begin{equation}
\label{eq:bias}
\mathrm{bias}(\theta^{(s)}) := \frac{1}{S} \sum^S_{s=1}\left( \theta^{(s)} \right)  - \theta,
\end{equation}

 \begin{equation}
\label{eq:rmse_s}
\mathrm{RMSE}(\theta^{(s)}) := \sqrt{ \frac{1}{S} \sum^S_{s=1} \left((\theta^{(s)}  - \theta)^2 \right) } = \sqrt{\mathrm{bias}(\theta^{(s)})^2 + \mathrm{Var} \left( \theta^{(s)} \right)},
\end{equation}
where $\mathrm{Var}(\theta^{(s)})$ denotes the variance over the posterior samples. Our analysis focuses on the true causal effect $\beta_{xy}$ and hence all metrics were computed for $\theta = \beta_{xy}$.
Furthermore, as we are interested in the size of the bias but not in its direction, we will present the absolute bias in our results. 

The above are reasonable measures for comparing models only if the assumed link coincides with the true link of the DGP. This is because the link determines the scale of the linear predictor and thus the comparability of the posterior samples $\{ \beta_{xy}^{(s)} \}$ with the true parameter value $\beta_{xy}$. 
To enable a comparison of models using different links, we also calculated the false positive rate (FPR; i.e., Type I-error rate) and the true positive rate (TPR; i.e., statistical power; inverse of the Type II-error rate) implied by the central 95\% credible interval of $\{ \beta_{xy}^{(s)} \}$. These metrics can be inferred from our simulations, because the true causal effect $\beta_{xy}$ was set to zero in some conditions (to study FPR) and to non-zero values in others (to study TPR).

For a deeper dive into posterior accuracy and calibration under model misspecification see \cite{scholz2023posterior}.

\subsubsection{Predictive performance}
In the absence of any case-specific arguments for a particular predictive metric, log-probability scores are recommended as a general-purpose choice \citep{vehtari_practical_2017}.
For this reason, we computed the expected log pointwise predictive density \citep[$\mathrm{ELPD}$;][]{vehtari_survey_2012, vehtari_practical_2017} as our main metric for predictive performance (operationalizing prediction), which is defined as
\begin{equation}
\label{eq:elpd}
    \text{ELPD}(y^*) := \sum_{i=1}^{N^*} \log p(y^*_i \mid \tilde{y}) = \sum_{i=1}^{N^*} \log \left(\frac{1}{S} \sum^S_{s=1} p(y^*_i \mgiven \theta^{(s)}) \right),
\end{equation}
where $\tilde{y}$ denotes the training data and $y^*$ denotes $N^* = 100$ independent test data points (previously unseen by the model) simulated from the same DGP configuration as the training data. 
We refer to the above metric as $\mathrm{ELPD}_{\rm test}$.
In addition, we also calculated $\mathrm{ELPD}$ via leave-one-out cross-validation (LOO-CV) as approximated via Pareto-smoothed importance sampling \citep[PSIS;][]{vehtari_practical_2017, Vehtari2022looR}.
We refer to this metric as $\mathrm{ELPD}_{\rm loo}$.
Approximate LOO-CV metrics have the advantage that they are readily and efficiently available also when analysing real data, at the expense of being only an approximation that might fail to estimate out-of-sample predictive performance accurately if there are influential observations \citep{vehtari_practical_2017}.
As we have the true data-generating process available during simulations, we used $\mathrm{ELPD}_{\rm test}$ to validate the results of $\mathrm{ELPD}_{\rm loo}$, but used the latter for our analyses as $\mathrm{ELPD}_{\rm loo}$ is available also in real-world scenarios.

In order to compare multiple models with respect to their ELPD performance,
we calculated the $\mathrm{ELPD}$ difference to the best performing model from a set $M := \{ m_1, ..., m_K \}$ of $K$ models all fit to the same training data:
\begin{equation}
\label{eq:elpd-delta}
    \Delta \mathrm{ELPD}(y^* \mid m_i) := \mathrm{ELPD}(y^* \mid m_i) - \text{max}(\mathrm{ELPD}(y^* \mid M)).
\end{equation}
As explained in Section \ref{sec:causal_foundation}, we always restricted comparisons (i.e., $M$) to models that used the same linear predictor. Additionally, for investigating posterior bias and RMSE performance, only models assuming the correct link function were compared.

\subsection{Statistical Analysis}
\label{sec:statistical-analysis}

To analyse the relationship between predictive performance, as measured by $\Delta \mathrm{ELPD}_{\rm loo}$ and $\Delta \mathrm{ELPD}_{\rm test}$, and parameter recoverability, as measured by absolute bias, RMSE, FPR, and TPR, we modeled the simulation results via Bayesian multilevel-models (BMMs).

Simulation results of models that did not converge were excluded from the analysis.
Specifically, we treated  models as converged if the posterior samples $\{ \beta_{xy}^{(s)} \}$ yielded $\widehat{R} < 1.01$ and $\mathrm{ESS} > 400$ \cite[for details on these thresholds, see][]{vehtari_rank-normalization_2021}. We further required models to have less than $10$ divergent transitions out of a total of 4000 post-warmup iterations. Ideally, we would like all models to converge with no divergent transitions, but this would have required extensive manual intervention to resolve all individual sampling problems, which is practically infeasible in our large simulation setup.
Within the set of convergent models, PSIS diagnostics indicated good $\mathrm{ELPD}_{\rm loo}$ approximations such that the latter can be treated as a trustworthy estimate of out-of-sample predictive performance \citep{vehtari_practical_2017}.
Finally, we set a lower threshold $\Delta \mathrm{ELPD}_{\rm loo} > - N \, (\, = -100)$ to exclude models that made exceptionally bad predictions compared to the best performing model on a given data set.
This resulted in apprimately 7\%
of models being filtered from the results.
Most of the filtered models (i.e., of the the approximately 7\%) from the double-bounded results were fit on data generated from a bathtub link (73\%) and filtered due to missing the $\Delta \mathrm{ELPD}_{\rm loo}$ threshold (94\%).
For the lower-bounded models, most filtered models failed to meet our convergence criteria (74\%) and were fit using a softplus link (73\%) and Fréchet likelihood.

In the linear predictor term of the BMMs, we included overall ('fixed') effects of $\Delta \mathrm{ELPD}$ in interaction with formula (i.e., the linear predictor term), (true) data generating link, and data generating shape, which we refer to as the \emph{global slopes} in the results in Section \ref{sec:results}.
Additionally, to account for the dependency of results obtained from the same simulated data set, we included varying ('random') effects across data sets of $\Delta \mathrm{ELPD}$ in interaction with formula which we refer to as the \emph{varying slopes} in the results in Section \ref{sec:results}. For the continuous positive metrics of parameter recovery, that is, absolute bias and RMSE, we assumed a log-normal likelihood, which enabled the use of highly optimized (log-)linear regression functions in Stan \citep{stan_2022}. For the binary metrics, that is, FPR and TPR, we assumed a canonical Bernoulli likelihood with logit link. The BMMs were estimated using a single MCMC chain with $500$ warmup and $1000$ post-warmup samples to keep the estimation time manageable (between one and three days wall-clock time per BMM).
All convergence metrics indicated sufficient convergence. As we were mainly interested in the qualitative patterns, rather than high resolution numerical results, we considered $1000$ post-warmup samples as sufficient, leading to effective sample sizes of a few hundred.

\section{Results}
\label{sec:results}

Recall, that this paper consists of two levels of Bayesian models. The first level are the models from the simulation study, that were fitted on simulated data as described in Section~\ref{sec:model-fitting}. The second level are meta-models that we fit on the metrics we extracted from the first level as described in Section~\ref{sec:statistical-analysis}. All results presented in this section are based on the meta-models to learn about trends across the entire simulation study.
Due to the vast number of simulation conditions, the main text illustrates selected results which are representative of the key patterns in the full simulation.
The full results are available in the supplementary material and in the online appendix \citep{online_appendix}.
In preliminary comparisons of $\mathrm{ELPD}_{\rm loo}$ and $\mathrm{ELPD}_{\rm test}$ results, we found both to be highly similar. Additionally, Pareto-$\hat{k}$ values of the $\mathrm{ELPD}_{\rm loo}$ approximation were generally low indicating good approximation accuracy \citep{vehtari_practical_2017}. For these reasons, we only present results of $\mathrm{ELPD}_{\rm loo}$ below due to its availability during real-world inference. 
Further, we do not present results for $| \mathrm{bias}(\beta_{xy}^{(s)}) |$ here, as they show the same qualitative trends as $\mathrm{RMSE}(\beta_{xy}^{(s)})$. Similarly, we will generally only show examples for either the double- or the lower-bounded data when the conclusions drawn from them match.
Some of the presented results show floor or ceiling effects, which point to a suboptimal hyperparameter calibration, discussed in more detail in Section \ref{sec:discussion}.

\begin{figure}
\centering
\includegraphics[width=0.99\linewidth]{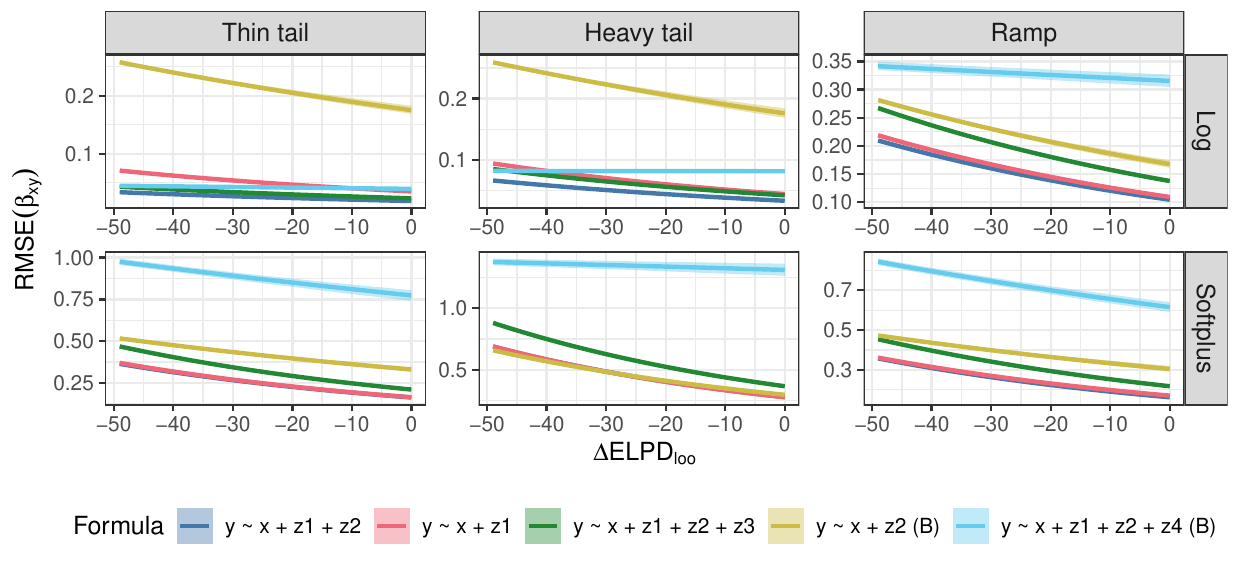}
\caption{Global slopes of $\Delta \mathrm{ELPD}_{\rm loo}$ on $\mathrm{RMSE}(\beta_{xy}^{(s)})$ for lower-bounded data and models. The shaded area represents the 95\% credible intervals of the posterior. The results are split by data generating link and shape as well as colored by formula, where (B) indicates the causally biased formulas. The x-axis is truncated at $-50$ to prevent unreasonable extrapolation to sparse data spaces. The negative slopes for causally unbiased formulas have fully negative 95\% credible intervals and indicate improved RMSE for better predicting models.}
\label{fig:positive-conditional-rmse-elpd}
\end{figure}

\begin{figure}
\centering
\includegraphics[width=0.99\linewidth]{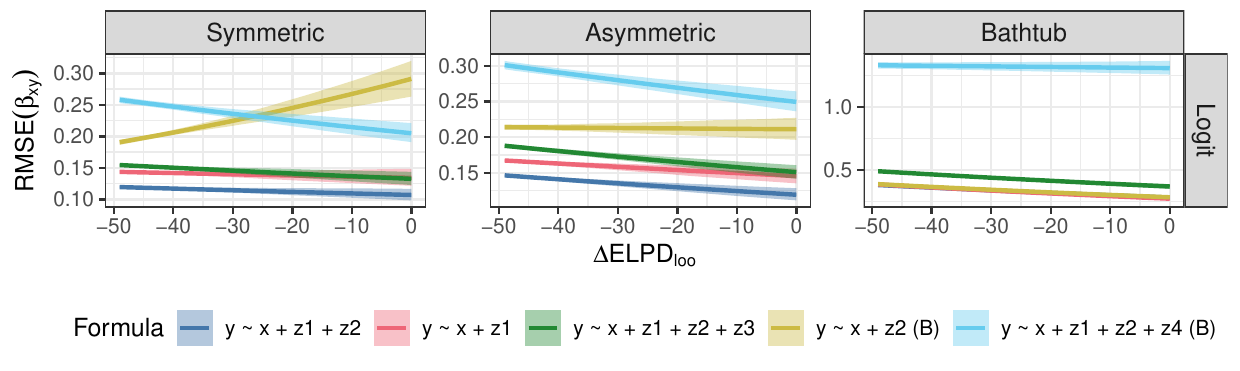}
\caption{Global slopes of of $\Delta \mathrm{ELPD}_{\rm loo}$ on $\mathrm{RMSE}(\beta_{xy}^{(s)})$ for double-bounded models and data generated with a logit link. The shaded area represents the 95\% credible intervals of the posterior. The results are split by data generating shape and colored by formula, where (B) indicates the causally biased formulas. The x-axis is truncated at $-50$ to prevent unreasonable extrapolation to sparse data spaces. The negative slopes for causally unbiased formulas have fully negative 95\% credible intervals and indicate improved RMSE for better predicting models.}
\label{fig:unit-conditional-rmse-elpd_logit}
\end{figure}

\subsection{RMSE}

Figure~\ref{fig:positive-conditional-rmse-elpd} displays the global slopes of $\Delta \mathrm{ELPD}_{\rm loo}$ on $\mathrm{RMSE}(\beta_{xy}^{(s)})$ for the lower-bounded data, grouped by formula (i.e., linear predictor), data generating link, and shape as explained in Section \ref{sec:statistical-analysis}.
The most important observation is that all slopes of unbiased formulas were negative, often very strongly so.
This means that, within a set of models fitted to the same data set and assuming the same causally unbiased formula, models that have higher (better) $\Delta \mathrm{ELPD}_{\rm loo}$ can be expected to have substantially lower (better) $\mathrm{RMSE}(\beta_{xy}^{(s)})$. While the exact size of the improvement is dependent on the simulation's hyperparameters, $\mathrm{RMSE}(\beta_{xy}^{(s)})$ reduces by around 50\% over the displayed span of $\Delta \mathrm{ELPD}_{\rm loo}$, an improvement of the same order of magnitude as the true effect $\beta_{xy}$.
Results of the double-bounded models showed the same behaviour.
For both data types, some of the unbiased formulas have rather flat slopes. This results from a floor effect in the $\mathrm{RMSE}$, as slopes for formulas that generally have a small error have little room to decrease further.
The causally biased formulas are less consistent, with some scenarios even showing positive slopes as exemplarily shown in Figure~\ref{fig:unit-conditional-rmse-elpd_logit} for double-bounded models fit on logit data. Additionally, even though the generally higher $\mathrm{RMSE}(\beta_{xy}^{(s)})$ of biased formulas would allow for steeper negative slopes, the biased slopes are often similar or even less negative than the unbiased ones.

In Table~\ref{tab:conditional_rmse_elpd}, we show the proportion of negative varying slopes across individual data sets. 
Similarly to the global slopes, the varying slopes of the unbiased formulas were largely negative (between 95\% and 99\% of the varying slopes were negative for the lower-bounded data and between 63\% and 94\% were negative for the double-bounded data). 
The exact proportion of negative slopes differed somewhat between ground-truth scenarios, as a result of specific hyperparameters choices in the calibration as well as random noise induced by the simulations.
However, the generally high proportions of negative varying slopes for the unbiased formulas indicates that the trend of lower (better) $\mathrm{RMSE}(\beta_{xy}^{(s)})$ for higher (better) $\Delta \mathrm{ELPD}_{\rm loo}$ is consistent across data sets.
For the causally biased formulas, the proportion of negative varying slopes were generally lower than for the unbiased formulas for each ground truth.
In addition to those overall trends, the proportion of negative slopes for the double-bounded models showed more variability between ground truths than the lower-bounded data. Especially the logit link had a smaller proportion than the other scenarios as a result of floor effects. 

\begin{threeparttable}
\caption{Proportion of negative varying slopes of $\Delta \mathrm{ELPD}_{\rm loo}$ on $\mathrm{RMSE}(\beta_{xy}^{(s)})$ for individual data sets (higher is better).}
\vspace*{1em}
\label{tab:conditional_rmse_elpd}
\begin{tabular}{lllllllll}
\hline
 & \multicolumn{3}{c}{Log} & \quad & \quad& \multicolumn{3}{c}{Softplus}\\
 & \multicolumn{1}{l}{\emph{Thin Tail}} & \multicolumn{1}{l}{\emph{Heavy Tail}} & \emph{Ramp} & \quad& \quad& \multicolumn{1}{l}{\emph{Thin Tail}} & \multicolumn{1}{l}{\emph{Heavy Tail}} & \emph{Ramp} \\ \hline
Unbiased   & \multicolumn{1}{l}{0.95}      & \multicolumn{1}{l}{0.96}       & 0.96 &\quad & \quad & \multicolumn{1}{l}{0.97}      & \multicolumn{1}{l}{0.99}       & 0.96 \\ 
Biased & \multicolumn{1}{l}{0.69}      & \multicolumn{1}{l}{0.59}       & 0.67 &\quad & \quad & \multicolumn{1}{l}{0.76}      & \multicolumn{1}{l}{0.71}        & 0.79 \\ \hline
\end{tabular}

\vspace*{1em}

\begin{tabular}{llllllllllll}
\hline
 & \multicolumn{3}{c}{Logit} & \quad & \multicolumn{3}{c}{Cauchit} & \quad & \multicolumn{3}{c}{Cloglog}\\
 & \multicolumn{1}{l}{\emph{Sym}} & \multicolumn{1}{l}{\emph{Asym}} & \emph{BT} & \quad& \multicolumn{1}{l}{\emph{Sym}} & \multicolumn{1}{l}{\emph{Asym}} & \emph{BT} & \quad& \multicolumn{1}{l}{\emph{Sym}} & \multicolumn{1}{l}{\emph{Asym}} & \emph{BT} \\ \hline
Unbiased   & \multicolumn{1}{l}{$0.63^*$}      & \multicolumn{1}{l}{0.69}       & 0.76 &\quad  & \multicolumn{1}{l}{0.9}      & \multicolumn{1}{l}{0.88}       & 0.91 &\quad & \multicolumn{1}{l}{0.83}      & \multicolumn{1}{l}{0.78}       & 0.94 \\ 
Biased & \multicolumn{1}{l}{0.51}      & \multicolumn{1}{l}{0.66}       & 0.64 &\quad & \multicolumn{1}{l}{0.9}      & \multicolumn{1}{l}{0.89}       & 0.92 &\quad & \multicolumn{1}{l}{0.73}      & \multicolumn{1}{l}{0.73}       & 0.93 \\ \hline
\end{tabular}
 \begin{tablenotes}
      \small
      \item $^*$ Floor effects in the global slopes were present for at least one of the respective formulas.
    \end{tablenotes}
\end{threeparttable}

\subsection{FPR and TPR}

\begin{figure}
\centering
\includegraphics[width=\linewidth]{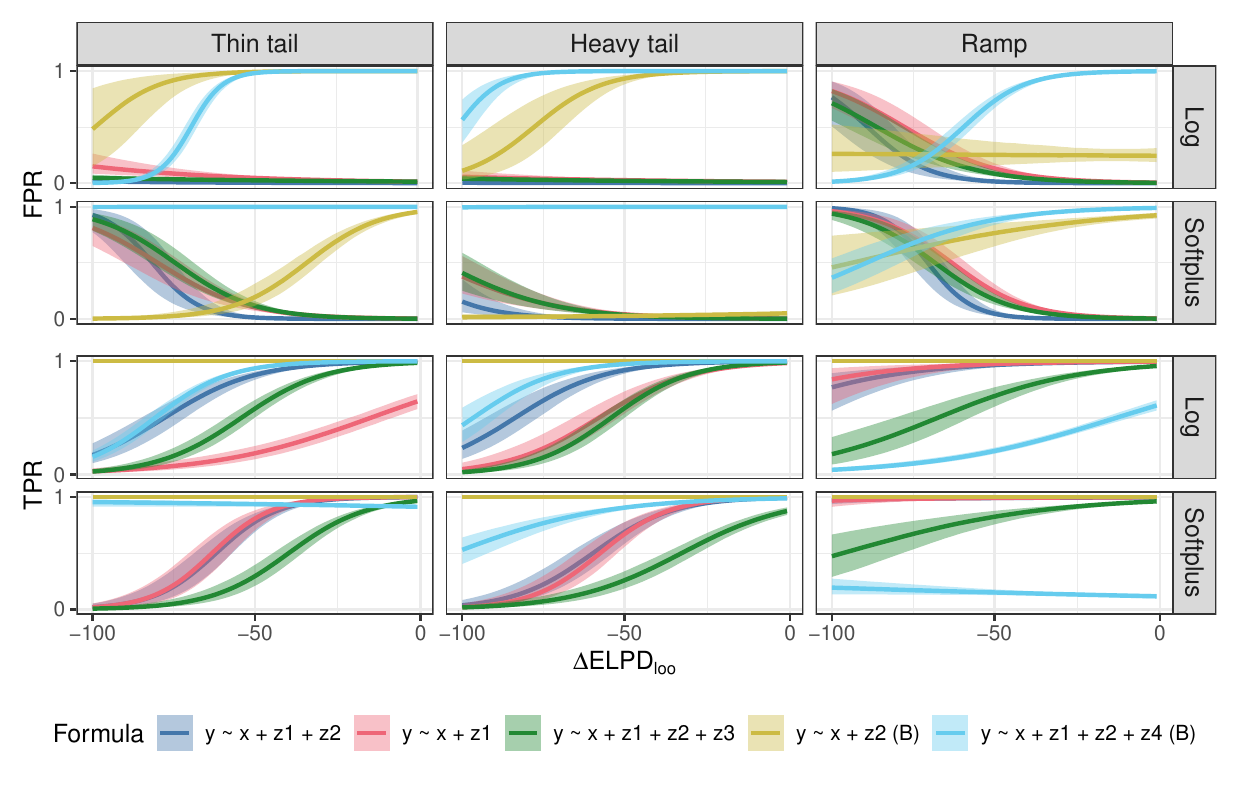}
\caption{Global slopes of $\Delta \mathrm{ELPD}_{\rm loo}$ on the false positive and true positive rates (FPR and TPR) for lower-bounded data. The shaded area represents the 95\% credible intervals of the posterior. The results are split by data generating link and shape as well as colored by formula, where (B) indicates the causally biased formulas. The low FPR and positive TPR slopes of the unbiased formulas indicate better Calibration for better predicting models.}
\label{fig:positive-tpr-fpr-elpd}
\end{figure}

Figure~\ref{fig:positive-tpr-fpr-elpd} shows the global slopes of $\Delta \mathrm{ELPD}_{\rm loo}$ on FPR and TPR for the lower-bounded data grouped by formula, data generating link, and shape as explained in Section \ref{sec:statistical-analysis}.
Starting with the FPR slopes, there is a clear distinction between the biased and unbiased formulas.
Where the unbiased formulas have negative slopes or even stay constant near $0.05$ (nominal Type-I error rate), the biased formulas have positive slopes or stay constant near 1 (i.e., $100$\% Type-I error rate) in all cases.
The results for the double-bounded scenarios show the same pattern. Acceptable FPR (i.e., values close to or less than the desired Type-I error rate) is a requirement for using TPR as a measure of parameter recovery, because it is trivial to achieve high TPR if FPR is allowed to be high at the same time. Accordingly, while the low FPR for the unbiased formulas implies the usefulness of TPR in these cases, the high FPR for the biased formulas makes their TPR results much less meaningful.

In the TPR results, we found a clear pattern of positive global slopes for the unbiased formulas whenever there are no ceiling effects (i.e., TPR = 1 regardless of $\Delta \mathrm{ELPD}_{\rm loo}$).
This indicates, that within models fitted to the same data set and using the same causally unbiased formula, models with higher (better) $\Delta \mathrm{ELPD}_{\rm loo}$ can be expected to have substantially higher (better) TPR.
The biased formulas again are less consistent in their slopes with cases of negative global slopes or shallower slopes even for generally low TPR, such that ceiling effects are not an issue.
Results of the double-bounded models showed the same behaviour with some additional cases of floor effects.

In Table~\ref{tab:tpr_fpr_slopes}, we show the proportion of positive varying slopes across individual data sets for the TPR.
For the causally unbiased formulas, the majority of the varying slopes are strongly positive, such that within a set of models fitted to the same data set and assuming the same causally unbiased formula, models that ranked higher in $\Delta \mathrm{ELPD}_{\rm loo}$ can be expected to have a higher TPR. This effect is rather consistent with positive proportions between $86\%$ and $95\%$ for different ground truths, besides the ramp and bathtub shape. The lower proportion of positive slopes for the ramp and bathtub shapes coincide with the floor and ceiling effects visible in Figure~\ref{fig:positive-tpr-fpr-elpd}. 

The biased formulas showed less consistency of the varying slope signs in the lower-bounded scenarios and similar consistency in the double-bounded scenarios.

\begin{threeparttable}
\caption{Proportion of positive varying slopes of $\Delta \mathrm{ELPD}_{\rm loo}$ on TPR (higher is better).}
\vspace*{1em}
\label{tab:tpr_fpr_slopes}
\begin{tabular}{lllllllll}
\hline
 & \multicolumn{3}{c}{Log} & \quad & \quad& \multicolumn{3}{c}{Softplus}\\
 & \multicolumn{1}{l}{\emph{Thin Tail}} & \multicolumn{1}{l}{\emph{Heavy Tail}} & \emph{Ramp} & \quad& \quad& \multicolumn{1}{l}{\emph{Thin Tail}} & \multicolumn{1}{l}{\emph{Heavy Tail}} & \emph{Ramp} \\ \hline
Unbiased   & \multicolumn{1}{l}{0.78}      & \multicolumn{1}{l}{0.86}       & 0.71 &\quad & \quad & \multicolumn{1}{l}{0.92}      & \multicolumn{1}{l}{0.87}       & $0.61^*$ \\ 
Biased & \multicolumn{1}{l}{$0.7^*$}      & \multicolumn{1}{l}{$0.59^*$}       & $0.51^*$ &\quad & \quad & \multicolumn{1}{l}{$0.54^*$}      & \multicolumn{1}{l}{$0.65^*$}        & $0.25^*$ \\ \hline
\end{tabular}

\vspace*{1em}

\begin{tabular}{llllllllllll}
\hline
 & \multicolumn{3}{c}{Logit} & \quad & \multicolumn{3}{c}{Cauchit} & \quad & \multicolumn{3}{c}{Cloglog}\\
 & \multicolumn{1}{l}{\emph{Sym}} & \multicolumn{1}{l}{\emph{Asym}} & \emph{BT} & \quad& \multicolumn{1}{l}{\emph{Sym}} & \multicolumn{1}{l}{\emph{Asym}} & \emph{BT} & \quad& \multicolumn{1}{l}{\emph{Sym}} & \multicolumn{1}{l}{\emph{Asym}} & \emph{BT} \\ \hline
Unbiased   & \multicolumn{1}{l}{0.94}      & \multicolumn{1}{l}{0.93}       & $0.24^*$ &\quad  & \multicolumn{1}{l}{0.95}      & \multicolumn{1}{l}{0.93}       & $0.78^*$ &\quad & \multicolumn{1}{l}{$0.86^*$}      & \multicolumn{1}{l}{0.94}       & $0.74^*$ \\ 
Biased & \multicolumn{1}{l}{$0.94^*$}      & \multicolumn{1}{l}{$0.94^*$}       & 0.81 &\quad & \multicolumn{1}{l}{$0.96^*$}      & \multicolumn{1}{l}{$0.88^*$}       & $0.87^*$ &\quad & \multicolumn{1}{l}{$0.48^*$}      & \multicolumn{1}{l}{$0.93^*$}       & 0.55 \\ \hline
\end{tabular}

 \begin{tablenotes}
      \small
      \item $^*$ Floor or ceiling effects in the global slopes were present for at least one of the respective formulas.
    \end{tablenotes}
\end{threeparttable}

\section{Discussion}
\label{sec:discussion}

In this paper, we studied the use of prediction as a proxy for explanation under several causal and statistical misspecification mechanisms in Bayesian GLMs.
We split our research question into two parts:
\begin{enumerate}[(i)]
    \item\label{enum:research-questions:PPPR2} Within a set of statistical models that all share the same underlying causal model, can prediction be reliably used as a proxy for explanation?
    \item Does the answer to (\ref{enum:research-questions:PPPR2}) depend on whether or not the causal model is biased?
\end{enumerate}

Within our results, we observed one consistent trend over almost all investigated scenarios:
When comparing statistical models sharing the same underlying \textit{unbiased} causal model, all considered measures of explanation (the causal parameter's absolute bias, RMSE, true positive rate and false positive rate) improved with improving out-of-sample prediction (measured by the expected log predictive density). 
Further, in almost all scenarios, the trends were also highly consistent across individual data sets, which indicates that the proxy can be reliably applied in practice, where inference usually concerns a single data set only.
In the few causally unbiased cases without a clear trend, the relationship was effectively zero due to floor and ceiling effects (further discussed below).
While better predicting models did not necessarily provide better explanation in those cases, prediction was (on average across datasets) not worsening explanation either.
For statistical models sharing the same underlying \textit{biased} causal model, the trends where inconsistent with examples for both improving and degrading explanation with improving prediction.
We conclude that, given a set of GLMs that all share the same unbiased causal model, prediction can be safely used as a proxy for explanation.

Our findings also offer empirical support for the more conceptual arguments of \cite{yarkoni_choosing_2017} and \cite{breiman_statistical_2001} that called for more attention to better predicting models, even if explanation is the main goal.
In that way, causality seems to be the missing link that connects the statistical relationship of explanation and prediction.

As is typical of simulation studies, their generalizability beyond the studied scenarios should be treated with caution.
While an analytic investigation supplementing the simulations would have been desirable, the lack of available closed-form posteriors outside of a few conjugate cases prevented such an investigation (as does the iterative nature of maximum likelihood in the models' frequentist counterparts).
We understand this paper as a promising starting point for further work on the relationship between explanation and prediction under different causal and statistical assumptions. Below, we offer our perspective on subsequent research questions.

As discussed in Section \ref{sec:glm}, we limited the scope of this paper to models with simple linear predictor terms.
We would expect to observe the same general trends for more complicated additive predictor terms where individual components may be non-linear, as the DAG-based misspecifications are agnostic to the details of the statistical model structure \citep{pearl_causal_2009}.
However, investigating the implications of (mis-)specifying individual terms of the linear predictor (e.g., modeling an effect as linear while it is truly non-linear in some way) might be interesting.
This would more strongly play into the question of overfitting in sparse scenarios with (relatively) small data yet complicated non-linear relationships to be approximated. Similar questions would arise when adding multilevel structure especially in sparse data scenarios where the (true) relevance of multilevel terms need to be balanced against their weak identification implied by the given data \citep{gelman2006data, buerkner_utility_2022}.

With regard to causal assumptions, we chose our data-generating processes (DGPs) and the four misspecified linear predictors in an attempt to span the most important and common classes of controls \citep{cinelli_crash_2020}.
Still, there remain many other possible DGPs and misspecifications, for example, unobserved variables or variables with measurement error \citep{kuroki2014measurement, tchetgen2020introduction}.
An extension of this work to different kinds of DGPs and misspecifications would ultimately add to the generalizability of results. That said, we currently do not see how an unbiased causal model would have to look like in order to render the relationship of prediction and explanation negative (i.e., better predicition implying worse explanation), as long as all compared models all shared the same unbiased causal model. 

In contrast, when comparing models with \textit{different} underlying causal models, the relationship between prediction and explanation is still not fully understood. Even though it is known, for example, that including colliders improves prediction but worsens explanation \citep{mcelreath_statistical_2020}, the situation for a set of different but all unbiased causal models has not yet been systematically explored to our knowledge. That is, when all compared statistical models have underlying \textit{unbiased} causal models, it is unclear if improved prediction implies improved explanation if the causal models are allowed to vary across the statistical models. We believe that this research direction would be one of the most important extensions of the present work.

In addition to the influence of the true and assumed causal models, we observed a noticeable effect of the true likelihood shapes assumed in our simulations.
The mechanisms by which different shapes imply different downstream results, including the extend of the relationship between prediction and explanation, remain somewhat unclear to us.
While the investigation of additional DGPs would generally offer potential for improving the generalizability of our findings, an extension of this work that would especially focus on different data-generating shapes and their properties more systematically could improve said understanding.
This directly ties into the problem of choosing reasonable ground-truth model parameters for the simulations. 
Fully Bayesian simulations tend to struggle when drawing parameters from prior distributions as this can easily lead to extreme and unrealistic data sets \citep{gabry_visualization_2019, mikkola_prior_2021}.
Our solution of carefully crafting scenarios that would represent specific, distinct, but still realistic, ground truths comes with the challenge of choosing an overwhelming amount of parameters. This did not fully succeed in all cases, as shown by the floor and ceiling effects that were present in a few scenarios, specifically for the true and false positive rates where causal relationships were sometimes too easy or too hard to detect.
While results influenced by those effects still support our overall conclusions, they reduce what we can learn from the affected scenarios.

Finally, in this study, we followed a Bayesian perspective on model building, despite using flat priors. The latter choice was primarily made to avoid some models having an 'unfair' advantage due to incidentally more suitable prior choices (see Section \ref{sec:model-fitting}). As a side effect of this choice, we think that our main results are likely to hold as well in case of frequentist (maximum likelihood) estimation of all models, and corresponding model-based metrics. 
We see it as unlikely that our conclusions would have changed in the light of applying (weakly-)informative priors \citep{stan_2022, mcelreath_statistical_2020, gelman_bayesian_2013}. However, there may be one subtle place where priors (or regularization more generally) may influence the relationship between prediction and explanation: Cross-validation measures of out-of-sample prediction produce correlated folds due to overlap in training data and have an intricate relationship with different true quantities they can be seen to approximate \citep{bates_cross-validation_2022}. These properties can be altered via regularization, especially in sparse regimes where the number of model parameters is high relative to the number of observations in the data \citep{bates_cross-validation_2022}. As such, the relationship between prediction and explanation may also vary in such scenarios, which would be interesting to study in the future.